\documentclass[showpacs,amssymb,pra,twocolumn]{revtex4}

\usepackage{graphicx}% Include figure files
\usepackage{dcolumn}% Align table columns on decimal point
\usepackage{bm}% bold math
\DeclareGraphicsExtensions{.eps}
\newcommand{\x}{\textbf{x}}
\newcommand{\q}{\textbf{q}}
\newcommand{\bV}[1]{\mbox{\boldmath${#1}$}}     % boldface

\begin{document}

\title{MULTIMODE SQUEEZING PROPERTIES OF A CONFOCAL OPO: \\
BEYOND THE THIN CRYSTAL APPROXIMATION}

\author{L.Lopez, S. Gigan, N.Treps, A. Ma\^\i tre, C. Fabre}

\affiliation{Laboratoire Kastler Brossel, Universit\'e Pierre et
Marie Curie, Campus Jussieu, Case 74, 75252 Paris cedex 05,
France}

\author{A. Gatti}

\affiliation{INFM, Dipartimento di Scienze Fisiche e Matematiche,
Universit\'a dell'Insubria, Via valleggio 11, 22100 Como, Italy}

\date{\today}
\begin{abstract}
Up to now, transverse quantum effects (usually labelled as
"quantum imaging" effects) which are generated by nonlinear
devices inserted in resonant optical cavities have been calculated
using the "thin crystal approximation", i.e. taking into account
the effect of diffraction only inside the empty part of the
cavity, and neglecting its effect in the nonlinear propagation
inside the nonlinear crystal. We introduce in the present paper a
theoretical method which is not restricted by this approximation.
It allows us in particular to treat configurations closer to the
actual experimental ones, where the crystal length is comparable
to the Rayleigh length of the cavity mode. We use this method in
the case of the confocal OPO, where the thin crystal approximation
predicts perfect squeezing on any area of the transverse plane,
whatever its size and shape. We find that there exists in this
case a "coherence length" which gives the minimum size of a
detector on which perfect squeezing can be observed, and which
gives therefore a limit to the improvement of optical resolution
that can be obtained using such devices.

\end{abstract}

\maketitle

%%%%%%%%%%%%%%%%%%%%%%%%%%%%%%%%%%%%%%%%%%%%%%%%%%%%%
\section{Introduction}
Nonlinear optical elements inserted in optical cavities have been
known for a long time to produce a great variety of interesting
physical effects, taking advantage of the field enhancement effect
and of the feedback provided by a resonant cavity
\cite{BoydBob,Siegman}. In particular, a great deal of attention
has been devoted to cavity-assisted nonlinear transverse effects,
such as pattern formation \cite{review1} and spatial soliton
generation \cite{review2}. More recently the quantum aspects of
these phenomena have begun to be studied, mainly at the
theoretical level, under the general name of "quantum imaging",
especially in planar or confocal cavities.

Almost all the investigations relative to intra-cavity nonlinear
effects, both at the classical and quantum level, have been
performed within the \textit{mean field approximation}, in which
one considers that the different interacting fields undergo only
weak changes through their propagation inside the cavity, in terms
of their longitudinal and transverse parameters. This almost
universal approach simplifies a great deal the theoretical
investigations, and numerical simulations are generally needed if
one wants to go beyond this approximation \cite{Leberre}. It
implies in particular that diffraction is assumed to be negligible
inside the nonlinear medium, which limits the applicability of the
method to nonlinear media whose length $l_{c}$ is much smaller
than the Rayleigh length $z_{R}$ of the cavity modes $z_{c}$ (so
called "thin" medium). This is a configuration that
experimentalists do not like much: they prefer to operate in the
case $l_{c} \simeq z_{c}$ which yields a much more efficient non
linear interaction for a given pump power \cite{boyd}. If one
wants to predict results of experiments in realistic situations,
one therefore needs to extend the theory beyond the usual thin
nonlinear medium approximation, and take into account diffraction
effects occurring together with the nonlinear interaction inside
the medium.

The effects of simultaneous diffraction and nonlinear propagation
have already been taken into account in the case of free
propagation, i.e. without optical cavity around the nonlinear
crystal, and they have been found to have a direct influence on
the shape of the propagating beam \cite{shen}. These effects have
also been studied in detail at the quantum level in the parametric
amplifier case \cite{Brambilla}, and recently for the soliton case
\cite{treps}. In contrast, they do not play a significant role
when the nonlinear medium is inserted in an optical cavity with
non degenerate transverse modes, which imposes the shape of the
mode. But they are of paramount importance in the case of cavities
having degenerate transverse modes, such as a plane or confocal
cavity, which do not impose the transverse structure of the
interacting fields, and which are used to generate multimode
quantum effects.

Within the thin crystal approximation, i.e. taking into account
diffraction effects only outside the crystal, striking quantum
properties have been predicted to occur in a degenerate OPO below
threshold using a confocal cavity \cite{Grangier,Petsas}: this
device generates quadrature squeezed light which is multimode in
the transverse domain. It was shown in the case of a plane pump
that the level of squeezing measured at the output of such an OPO
neither depends on the spatial profile of the local oscillator
used to probe it, nor on the size of the detection region. This
implies that a significant quantum noise reduction, in principle
tending to perfection when one approaches the oscillation
threshold from below, can be observed in arbitrarily small
portions of the down-converted beam. Therefore in this model there
is no limitation in the transverse size of the domains in which
the quantum noise is reduced when the OPO works in the exact
confocal configuration. Such a multimode squeezed light appears
thereby as a very promising tool to increase the resolution in
optical images beyond the wavelength limit.

It is therefore very important to make a more realistic
theoretical model of this system, which is no longer limited by
the thin crystal approximation, to see whether the predicted local
squeezing is still present in actual experimental realizations in
which the crystal length is of the order of the Rayleigh range of
the resonator. This is the purpose of the present paper, in which
we will show that the presence of a long crystal inside the
resonator imposes a lower limit to the size of the regions in
which squeezing can be measured ("coherence area"), which is
proportional to $w_c^2 l_c/z_R$, where $w_c$ is the cavity beam
waist, $l_c$ is the crystal length and $z_R$ the Rayleigh range of
the resonator.

The following section (section II) is devoted to the general
description of the model that is used to treat the effect of
diffraction inside the crystal, using the assumption that the
single pass nonlinear interaction is weak in the crystal. We then
describe in section III the method that is used to determine the
squeezing spectra measured in well-defined homodyne detection
schemes. We give in sections IV and V the results for such
quantities respectively in the near field and in the far field,
and conclude in section VI.

\section{The model}
\subsection{Assumptions of the model}
\begin{figure}[ht]
\centerline{
    \scalebox{.55}{\includegraphics*{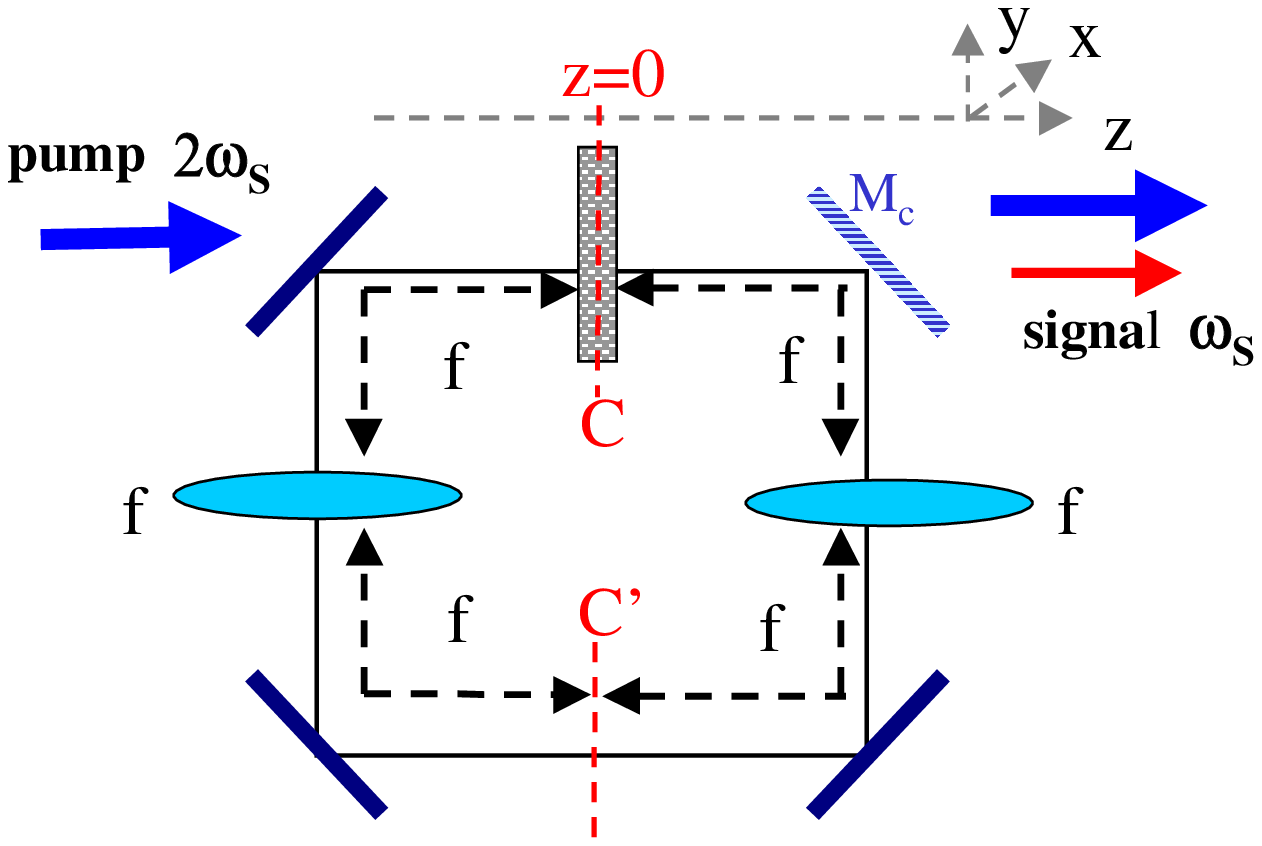}} }
\caption{Confocal ring cavity. The mirrors transmit the pump wave
and reflect the signal wave, with the exception of mirror $M_c$
that partially transmits the signal.} \label{fig:fig1}
\end{figure}

Let us consider a confocal cavity, that  for simplicity we take as
a ring cavity of the kind shown schematically in Fig.
\ref{fig:fig1} (\cite{Schwob}\cite{Mancini}).
It is formed by four plane mirrors and two lenses having a focal
length equal to one quarter of the total cavity length, and
symmetrically placed along the cavity, so that the focal points
coincides at two positions C and C'. It contains a type I
parametric medium of length $l_c$, centered on the  point C (see
figure). It is pumped by a field $A_p$ of frequency $2\omega_s$
having a Gaussian shape and focused in the plane containing the
point C. In such a plane the variation of the mean envelope with
the transverse coordinate \textbf{x} is given by :
\begin{equation}
  A_{p}(\textbf{x}) = A_p exp(-\frac{|\textbf{x}|^2}{w_p^2})
\label{pump}
\end{equation}
We assume that the mirrors are totally transparent for the pump
wave, and perfectly reflecting for the field at frequency
$\omega_s$, except for the coupling mirror $M_c$, which has a
small transmission $t$ at this frequency. The system was described
in \cite{Petsas} under the thin parametric medium approximation.
We will follow here the same approach, generalized to the case of
a thick parametric medium of length $l_c$. The intracavity signal
field at frequency $\omega_s$ is described by a field envelope
operator $\hat{B}(\textbf{x},z)$, where $z$ is the longitudinal
coordinate along the cavity  ($z=0$ corresponding to plane C),
obeying the standard equal time commutation relation at a given
transverse plane at position $z$:
\begin{equation}
  [\hat{B}(\textbf{x},z,t),\hat{B}^\dagger (\textbf{x'},z,t)]=\delta(\textbf{x}-\textbf{x'}) .
\end{equation}
As we are only interested in the regime below threshold and
without pump depletion, the pump field fluctuations do not play
any role.

In a confocal resonator the cavity resonances corresponds to
complete sets of Gauss-Laguerre modes with a given parity for
transverse coordinate inversion; we assume that a set of cavity
\textit{even} modes is tuned to resonance with the signal field,
and that the odd modes are far off-resonance. It is then useful to
introduce the even part of the field operator:
\begin{equation}
  \hat{B}_{+}(\textbf{x},z,t)=\frac{1}{2}\{\hat{B}(\textbf{x},z,t)+\hat{B}(-\textbf{x},z,t)]\, ,
\label{B+def}
\end{equation}
which obeys a modified commutation relation:
\begin{equation}
  [\hat{B}_{+}(\textbf{x},z,t),\hat{B}_{+}^\dagger (\textbf{x'},z,t)]=\frac{1}{2}[\delta(\textbf{x}-\textbf{x'})+\delta(\textbf{x}+\textbf{x'})]
\label{commutatr}
\end{equation}
and can be written as an expansion over the even Gauss-Laguerre
modes:
\begin{equation}
  \hat{B}_{+}(\textbf{x},z,t)=\sum_{p,l even}
  f_{p,l}(\textbf{x},z)\hat{a}_{p,l}(z,t)\, ,
\label{expansion}
\end{equation}
where $\hat{a}_{p,l}(z,t)$ is the annihilation operator of a
photon in mode $(p,l)$ at the cavity position $z$ and at time $t$.

The interaction Hamiltonian of the system in the interaction
picture is given by
\begin{equation}
H_{int}=\frac{i\hbar g}{2 l_c}\int_{-l_c/2}^{l_c/2}dz'\int\int d^2
x'
\{A_P(\textbf{x'},z')[\hat{B_{+}^{\dagger}}(\textbf{x'},z',t)]^2-
h.c.\} \label{Hamiltonian} \, ,
\end{equation}
where g is the coupling constant proportional to the second order
nonlinear susceptibility $\chi^{(2)}$. This equation generalizes
the thin medium parametric Hamiltonian of Ref.\cite{GattiLugiato}.

\subsection{Evolution equation in the image plane (near-field)}
In previous approaches \cite{Grangier,Petsas}, the crystal was
assumed to be thin, so that one could neglect the longitudinal
dependence of $A_P$ and $\hat{B}_{+}$ along the crystal length in
the Hamiltonian (\ref{Hamiltonian}). This cannot be done in a
thick crystal. We will nevertheless make a simplifying assumption
which turns out to be very realistic in the c.w. regime, with pump
powers below 1W. We assume that the nonlinear interaction is very
weak, so that it does not affect much the field amplitudes in a
single pass through the crystal. We will therefore remove the $z$
dependence of the operators $\hat{a}_{p,l}$ in
Eqs.(\ref{expansion}, \ref{Hamiltonian}), assuming
 $\hat{a}_{p,l}(z,t)=\hat{a}_{p,l}(z=0,t)= \hat{a}_{p,l}(t)$, where $z=0$ is the crystal/cavity center C. The longitudinal variation of the signal
operator $\hat{B}$ is then only due to diffraction and is
described by the well-known $z$ dependence of the modal functions
$f_{p,l}(\textbf{x},z)$. This assumption leads to a rather simple
expression of the commutator for the $\hat{B}_{+}$ field at
different positions inside the crystal:
\begin{equation}
 [\hat{B}_{+}(\textbf{x},z,t),\hat{B}_{+}^\dagger (\textbf{x'},z',t')]=
 G_+^*(z-z' ; \textbf{x},\textbf{x'}) \, .
\end{equation}
Here  $G_+(z; \textbf{x},\textbf{x'})$ is the symmetrized part of
the Fresnel propagator $G(z ; \textbf{x},\textbf{x'})$, describing
the field linear propagation inside the crystal:
\begin{equation}
G_+(z; \textbf{x},\textbf{x'})=\frac{1}{2}[G(z ;
\textbf{x},\textbf{x'})+G(z ; \textbf{x},-\textbf{x'})]\, .
 \end{equation}
with
\begin{equation}
\label{Fresnel} G(z ; \textbf{x},\textbf{x'})=\frac{i k_s}{2\pi z}
e^{  i k_s \frac{ |  \x  -\x' - \bV{\rho}_s z  |^2}{2z}} \, ,
 \end{equation}
where $k_s= n_s \omega_s /c$ is the field wavenumber, with $n_s$
being the index of refraction at frequency $\omega_s$, and we have
introduced a walk-off term, present only if the signal wave is an
extraordinary one, described by the two-dimensional walk-off angle
$\bV{\rho}_s$.

It is now possible to derive the time evolution of the field
operator $\hat{B}(\textbf{x},z,t)$ due to the parametric
interaction. We will for example calculate it at the mid-point
plane $z=0$ of the crystal :
\begin{equation}
\left. \frac{\partial \hat{B}_{+}}{\partial
t}(\textbf{x},0,t)\right|_{int}=g\int\int d^2 x"
K_{int}(\textbf{x},\textbf{x"})\hat{B}_{+}^{\dagger}(\textbf{x"},0,t)
\end{equation}
with the integral kernel $K_{int}$ given by :
\begin{eqnarray}
\label{thick}
K_{int}(\textbf{x},\textbf{x"})&=&\frac{1}{l_c}\int_{-l_c/2}^{l_c/2}dz'\int\int d^2 x'\\
& & A_P(\textbf{x'},z')G_+^*(z';
\textbf{x'},\textbf{x})G_+^*(z'; \textbf{x'},\textbf{x"})\, .\nonumber
\end{eqnarray}
In the limit of a thin crystal considered in Refs.\cite{Grangier,
Petsas}, Eq. (\ref{thick}) is replaced by the simpler expression :
\begin{equation}
\label{thin} \frac{\partial \hat{B}_{+}}{\partial
t}(\textbf{x},0,t)|_{int}=g A_P(\textbf{x})
\hat{B}_{+}^{\dagger}(\textbf{x},0,t)
\end{equation}
In the thin crystal case (Eq.(\ref{thin})), the parametric
interaction is  local, i.e. the operators at different positions
of the transverse plane are not coupled to each other, whereas in
the thick crystal case (Eq.(\ref{thick})), the parametric
interaction mixes the operators at different points of the
transverse plane, over areas of finite extension. Note however
that operators corresponding to different $z$ values are not
coupled to each other, because of our assumption of weak
parametric interaction. This situation is very close to the one
considered in refs.\cite{Kolobov1,Kolobov2,Brambilla} for
parametric down-conversion and amplification in a single-pass
crystal, where finite transverse coherence areas for the spatial
quantum effects arise because of the finite spatial emission
bandwidth of the crystal. In a similar way, in our case the
spatial extension of the kernel $K_{int}$ will turn out to give
the minimum size in which spatial correlation or local squeezing
can be observed in such a system. The analogy will become more
evident in the next section, where we will explicitly solve the
propagation equation of the Fourier spatial modes along the
crystal.

In order to get the complete evolution equation for the signal
beam, one must add the free Hamiltonian evolution of the
intracavity beam and the damping effects. This part of the
treatment is standard \cite{Gardiner}, and is identical to the
case of a thin crystal inserted in a confocal cavity
\cite{Petsas}. The final evolution equation reads:
\begin{eqnarray}
&&\frac{\partial \hat{B}_{+}}{\partial
t}(\textbf{x},0,t)=-\gamma(1+i\Delta)\hat{B}_{+}(\textbf{x},0,t) \\
&&+ g \int\int d^2 x"
K_{int}(\textbf{x},\textbf{x"})\hat{B}_{+}^{\dagger}(\textbf{x"},0,t)+\sqrt{2
\gamma}\hat{B}_{+ \, in}(\textbf{x},0,t)\nonumber \label{evolution}
\end{eqnarray}
where $\gamma$ is the cavity escape rate, $\Delta$ the normalized
cavity detuning of the even family of modes closest to resonance
with the signal field, and $\hat{B}_{in}$ the input field
operator.

\begin{figure}[ht]
\centerline{
    \scalebox{.45}{\includegraphics*{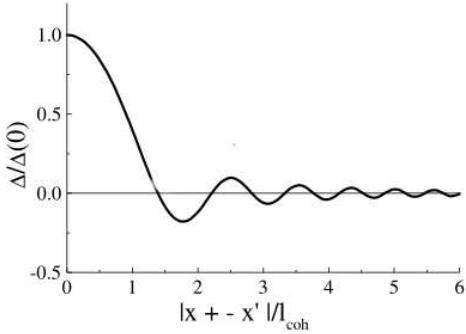}} }
\caption{Evaluation of the coupling kernel.  $\Delta$ given by Eq.
(\ref{delta1}) is plotted as a function of $|\textbf{x} \pm
\textbf{x"}|$ scaled to the coherence length (\ref{lcoh}). The
first zero of $\Delta$ is obtained for the value $1.37$ of the
coordinate.} \label{fig:fig2}
\end{figure}

%\subsubsection{Evaluation of the coupling kernel}
In order to evaluate the coupling kernel, let us first take into
account the diffraction of the pump field, focussed at the center
of the crystal, $z=0$. It is described by the Fresnel propagator
$G_p(z ; \textbf{x},\textbf{x'})$, equal to (\ref{Fresnel}) when
one replaces $k_s$ by the pump wavenumber $k_p$, and the signal
walk-off angle $\bV{\rho}_s$ with the pump walk-off angle
$\bV{\rho}_p$.  One then gets :
\begin{eqnarray}
&&K_{int}(\textbf{x},\textbf{x"})=\frac{1}{l_c}\int_{-l_c/2}^{l_c/2}dz'
\int d^2 x' \int d^2 y  \\
&&A_p(\textbf{y}) G_p(z',0 ;
\textbf{x'},\textbf{y}) G_+^*(z',0 ;
\textbf{x'},\textbf{x})G_+^*(z',0; \textbf{x'},\textbf{x"})\nonumber
\end{eqnarray}
Assuming for simplicity  exact collinear phase matching
$k_p=2k_s$, and neglecting the walk-off of the extraordinary wave
, four of the five integrations can be exactly performed, and one
finally gets:
\begin{eqnarray}
K_{int}(\textbf{x},\textbf{x"}) &=&\frac{1}{2}\left[
A_p(\frac{\textbf{x}+\textbf{x"}}{2})
\Delta(\textbf{x}-\textbf{x"})\right.\nonumber\\
&&\left. +A_p(\frac{\textbf{x}-\textbf{x"}}{2})\Delta(
\textbf{x}+\textbf{x"}  )   \right] \label{Kint}
\end{eqnarray}
with
\begin{equation}
\Delta( \textbf{x} \pm \textbf{x"} )=\frac{ik_s}{4 \pi l_c}
\int_{-l_c/2}^{l_c/2}\frac{dz'}{z'} e^{\frac{ik_s}{4z'}|\textbf{x}
\pm \textbf{x"} |^2} \label{delta1}
\end{equation}
It can be easily shown that the function $\Delta( \textbf{x} \pm
\textbf{x"} )$ tends to the usual two-dimensional distribution
$\delta(\textbf{x} \pm \textbf{x"} )$ when $ l_c\rightarrow 0$,
and that it can  be written in terms of the integral sine function
$Si(x)=\int_{0}^{x}\frac{sinudu}{u}$ \cite{Abramowitz}
\begin{equation}
\Delta( \textbf{x} \pm \textbf{x"} )=\frac{k_s}{2 \pi l_c}\left(
\frac{\pi}{2} -Si(\frac{k_s|\textbf{x} \pm \textbf{x"} |^2}{2
l_c}) \right)
\end{equation}
This expression shows us that $\Delta$ takes negligible values
when $|\textbf{x} \pm \textbf{x"} |\gg \sqrt{\frac{\lambda
l_c}{\pi n_s}}$. Fig.\ref{fig:fig2} plots $\Delta$ as a function
of the distance $|\textbf{x} \pm \textbf{x"} |$ scaled to
\begin{equation}
l_{coh}=\sqrt{\frac{\lambda l_c} {\pi n_s}}= w_C
\sqrt{\frac{l_c}{n_s z_C}}\, , \label{lcoh}
\end{equation}
where $w_C$ and $z_C$ are the cavity waist and Rayleigh range,
respectively.
%%%%FIGURA KERNEL %%%%%%%%
This expression shows that when the crystal length is on the order
of the Rayleigh range of the resonator, the transverse coherence
length is on the order of the cavity waist. Recalling that the
pump field has a Gaussian shape of waist $w_p$, in order to have a
multimode operation one must therefore use a defocussed pump, with
$w_p \gg w_c$, or alternatively use a crystal much shorter than
the Rayleigh range of the resonator, which is detrimental for the
oscillation threshold of the OPO. The relevant scaling parameter
of our problem is therefore
\begin{equation}
b=\frac{ w_p^2}{l_{coh}^2}= 2 n_s \frac{z_p}{l_c} \, , \label{b}
\end{equation}
where $z_p$ is the Rayleigh or diffraction length of the pump
beam. This parameter sets the number of spatial modes that can be
independently excited, and it will turn out to give also the
number of modes that can be independently squeezed.

\subsection{Evolution equation in the spatial Fourier domain (far-field)}
In this section we will investigate the intracavity dynamics of
the spatial Fourier amplitude of the signal field, which will
offer an alternative formulation of the problem. Fourier modes can
be observed in the far-field plane with respect to the crystal
center C, which in turn can be detected in the focal plane of a
lens placed outside the cavity. Let us introduce the spatial
Fourier transform of the signal field envelope operator
\begin{eqnarray}
\hat{B}_{+}(\textbf{q},z,t)&=&
\int\frac{d^{2}x}{2\pi}\hat{B}_{+}(\textbf{x},z,t)e^{-i\textbf{q}\cdot\textbf{x}}\nonumber \\
&=& \frac{1}{2} \left[\hat{B}(\textbf{q},z,t)
+\hat{B}(-\textbf{q},z,t) \right]
\end{eqnarray}
Equation (\ref{evolution}) becomes:
\begin{eqnarray}
&&\frac{\partial \hat{B}_{+}}{\partial
t}(\textbf{q},0,t)=-\gamma(1+i\Delta)\hat{B}_{+}(\textbf{q},0,t)+ \\
&& g \int d^2 q"
\tilde{K}_{int}(\textbf{q},\textbf{q"})\hat{B}_{+}^{\dagger}(\textbf{q"},0,t)+\sqrt{2
\gamma}\tilde{B}_{+ \, in}(\textbf{q},0,t) \, , \nonumber \label{evolution2}
\end{eqnarray}
where the  coupling Kernel
$\tilde{K}_{int}(\textbf{q},\textbf{q"})$ is the Fourier transform
of the kernel (\ref{Kint}) with respect to both arguments.
Straightforward calculations show that
\begin{eqnarray}
\tilde{K}_{int}(\textbf{q},\textbf{q'})&=&\frac{1 }{2}\left(
\tilde{A_{p}}(\textbf{q}+\textbf{q'}) {\rm
sinc}[\frac{l_{c}}{2k_{s}}|\frac{\textbf{q}-\textbf{q'}}{2}|^2]\right.\nonumber\\
&&\left.+\tilde{A_{p}}(\textbf{q}-\textbf{q'})
{\rm
sinc}[\frac{l_{c}}{2k_{s}}|\frac{\textbf{q}+\textbf{q'}}{2}|^2
]\right) \, , \label{Ktilde}
\end{eqnarray}
where $\tilde{A_p}$ is the spatial Fourier transform of the
Gaussian pump profile (\ref{pump}), i.e. $
\tilde{A}_{p}(\textbf{q}) = \frac{w_p^2}{2}A_p
\exp{(-|\textbf{q}|^2 \frac{ w_p^2}{4 })} $.

The result (\ref{Ktilde}) can  be also  derived by solving the
propagation equation of the pump and signal wave inside a
$\chi^{(2)}$ crystal directly in the Fourier domain and in the
limit of weak parametric gain. We will follow here the same
approach as in \cite{Brambilla} and \cite{GattiStokes}, and write
the propagation equation in terms of the spatio-temporal Fourier
transform field operators $\hat{A}_j (\q, \omega, z)$ of the pump
($j=p$) and signal ($j=s$) waves. Since the cavity linewidth is
smaller by several orders of magnitude than the typical frequency
bandwidth of the crystal, the cavity filters a very small
frequency bandwidth around the carrier frequency $\omega_s$ of the
signal; moreover, we have assumed that the pump is monochromatic,
so that we can safely neglect the frequency argument in the
propagation equations, which take the form
\begin{equation}
\frac{\partial \hat{A}_{j}}{\partial z} (\textbf{q},z)  =i k_{jz}
(\q) \hat{A}_j (\textbf{q},z) + \hat{P}^{NL}_j(\textbf{q},z) \, ,
\label{ddz1old}
\end{equation}
where $\hat{P}^{NL}_j$ is the nonlinear term, arising from the
second order nonlinear susceptibility of the crystal. $k_{jz} (\q)
= \sqrt{k_j^2 -q^2}$ is the projection along the z-axis of the
wavevector, with $k_j= k_j( \omega_j, \q)$ being the wave-number,
which for extraordinary waves depends also on the propagation
direction (identified by $\q$). For the pump wave, we assume an
intense coherent beam, that we suppose undepleted by the
parametric down-conversion process in a single pass through the
crystal, so that
\begin{equation}
\hat{A}_p (\textbf{q},z) \to \tilde{A}_p(\textbf{q},z)  = e^{i
k_{pz} (\q) z}A_p(\textbf{q},0)  \, ,
\end{equation}
where we take the crystal center as the reference plane $z=0$. For
the signal, the propagation equation is more easily solved by
setting $ \hat{A}_s(\textbf{q},z)  = \exp{ (i k_{sz} (\q)
z})\hat{a}_s (\textbf{q},z)$. The evolution along $z$ of the
operator $\hat{a}_s$ is only due to the parametric interaction and
is governed by the equation (see e.g.\cite{Brambilla} and
\cite{GattiStokes} for more details):
\begin{equation}
\frac{\partial \hat{a}_{s}}{\partial z} (\textbf{q},z)  =
\frac{\sigma}{l_c} \int d^2 q' \,A_p  (\q+\q',0)
\hat{a}_s^{\dagger} (\textbf{q}',z) e^{i \delta (\q,\q') z} \, ,
\label{ddz1}
\end{equation}
where  $\sigma/l_c$ is the parametric gain per unit length, and we
have introduced the phase mismatch function
\begin{equation}
\delta(\q,\q') = k_{pz} (\q+\q')   -k_{sz} (\q)  -k_{sz}(\q').
\label{delta}
\end{equation}
Equation (\ref{ddz1}) has the formal solution
\begin{eqnarray}
\hat{a}_{s} (\textbf{q},\frac{l_c}{2}) & = &\hat{a}_{s} (\textbf{q},
-\frac{l_c}{2}  )
 + \frac{\sigma}{l_c}  \int_{-\frac{l_c}{2}
}^{\frac{l_c}{2}} dz' \int d^2 q'\nonumber \\
&& A_p  (\q+\q',0)
\hat{a}_s^{\dagger} (\textbf{q}',z') e^{i \delta (\q,\q') z'} \, .
\label{ddz2old}
\end{eqnarray}
Assuming a weak parametric efficiency $\sigma \ll 1$, we can solve
this equation iteratively. At first order in $\sigma$ the solution
reads:
\begin{equation}
\hat{a}_{s} (\textbf{q},\frac{l_c}{2})  = \hat{a}_{s} (\textbf{q},
-\frac{l_c}{2}  )  + \sigma  \int d^2 q' \, K_1 (\q, \q')
\hat{a}_s^{\dagger} (\textbf{q}',0) \, , \label{ddz2}
\end{equation}
with
\begin{equation}
K_1 (\q,\q') = \tilde{A}_p  (\q+\q',0) \rm{sinc} {\left[\delta
(\q,\q') \frac{l_c}{2}\right] } \, . \label{K1}
\end{equation}
We observe that in the paraxial approximation $k_{jz}(\q) \approx
k_j -{\bf\rho}_j \cdot \q - \frac{q^2}{2k_j}$, where ${\bf
\rho}_j$ is the  walk-off angle and $k_j=n_j \omega_j/c$. The
phase mismatch function is hence given by:
\begin{eqnarray}
\delta(\q, \q') &=& k_p -2k_s  + ( {\bf \rho}_s -{\bf \rho}_p) \cdot
(\q +\q')\nonumber \\
&& - \frac{|\q + \q'|^2}{2k_p} + \frac{1}{2k_s} (q^2 +
q'^2)
\end{eqnarray}
Assuming exact phase matching $k_p=2k_s$, and neglecting the
walk-off term, the argument of the sinc function in Eq. (\ref{K1})
becomes
\begin{equation}
\delta(\q, \q') \frac{l_c} {2} = \frac{l_c}{2k_s}  \left|\frac{\q
- \q'}{2}\right|^2
\end{equation}
In this way we start to recover the result of the Hamiltonian
formalism used to derive Eqs.(\ref{evolution2}) and
(\ref{evolution}), where, however, the effect of walk-off and
phase mismatch were neglected for simplicity. Indeed, it is not
difficult to show that the variation of the intracavity field
operator $ \hat{B}_+ (\q,0,t) $ per cavity round trip time $\tau$,
due to the parametric interaction in a single pass through the
crystal is
\begin{eqnarray}
 \left. \frac{1}{\tau}\Delta \hat{B}_+ (\q,0,t) \right|_{int} & = & \frac{\sigma}{\tau}  \int d^2 q' \, \frac{1}{2} \left[K_1 (\q, \q')\right. \nonumber\\
& &\left. + K_1 (\q, -\q')\right] \hat{B}_+^{\dagger} (\textbf{q}',0,t) \, .
\end{eqnarray}
This approach permits us to understand the physical origin of the
sinc terms in the coupling kernel of Eq. (\ref{Ktilde}) (which are
the Fourier transform of the $\Delta$ terms in Eq. (\ref{Kint})),
that is the limited phase-matching bandwidth of the nonlinear
crystal. For a crystal of negligible length, phase matching is
irrelevant and there is no limitation in the spatial bandwidth of
down-converted modes, whereas for a finite crystal the cone of
parametric fluorescence has an aperture limited to a bandwidth of
transverse wavevectors $\Delta q \approx 1/l_{coh} \propto 1/
\sqrt{\lambda l_c}$. As a consequence of the confocal geometry,
the cavity ideally transmits  all the Fourier modes, so that the
only limitation in spatial bandwidth is that arising from phase
matching along the crystal.

We notice that if the pump is defocussed enough, the
phase-matching limitation results in a limitation of the spot size
$\propto 1/l_{coh} $ in the far-field with respect to the cavity
center. Inside this spot, modes are coupled because of the finite
size of the pump beam (the terms $\propto \tilde{A_P}$ in
Eq.(\ref{Ktilde})), inside a region of size $\propto w_p^{-1}$.
The relevant parameter which  sets the number of Fourier modes
that can be independently excited is again given by
$b=w_p^2/l_{coh}^2$ (see Eq. (\ref{b})).

\section{Homodyne detection and squeezing spectrum}

\subsection{Homodyne detection scheme in the far field and near field}
\begin{figure}[ht]
\centerline{
    \scalebox{.35}{\includegraphics*{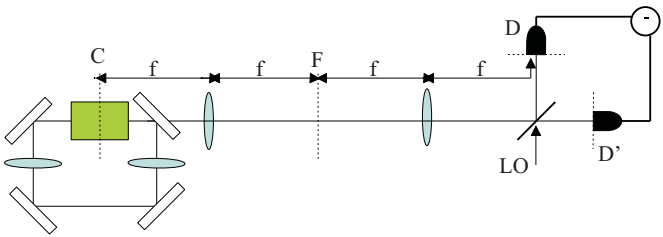}} }
\caption{Balanced homodyne detection scheme in the near field. Two
matching lenses of focal f are used to image the cavity center C
at the detection planes D and D'} \label{fig:fig3}
\end{figure}

 The method  used for measuring the noise-spectrum outside the cavity
 is a balanced homodyne detection scheme \cite{Special}. We will use
two configurations: the near-field configuration (x-position basis
described in II.B) and the far-field configuration (q-vector basis
described in II.C). The complete detection scheme in the
near-field case is schematically shown in Fig. \ref{fig:fig3}. The
two matching lenses of focal length \textit{f} image the
crystal/cavity center plane C onto the detection planes D and D'.
The image focal plane F of the first lens coincides with the
 object focal plane of the second one, and represents the far-field plane
  with respect to the cavity center C. In planes C,F,D the signal field
has its minimum waist, and it has a flat wavefront.

\begin{figure}[ht]
\centerline{
    \scalebox{.35}{\includegraphics*{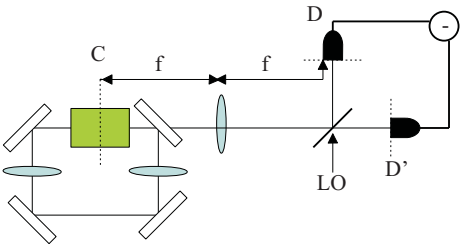}} }
\caption{Balanced homodyne detection scheme in the far field. A
matching lens of focal f is used to make the far field image of
the cavity center C at the detection planes D and D'}
\label{fig:fig4}
\end{figure}

 The detection scheme in the far field is obtained by using only one lens
 as depicted in Fig. \ref{fig:fig4}. The focal length \textit{f} lens is
 used to image the far field plane with respect of the cavity center C
 onto the detection plane D.

The symmetrical beam-splitter \textit{BS} (reflection and
transmission coefficients $\textit{r}=\frac{1}{\sqrt{2}}$ and
$\textit{t}=\frac{1}{\sqrt{2}}$) mixes the output signal field
with an intense stationary and coherent beam
$\alpha_{L}(\textbf{x},z)$, called local oscillator (LO). Note
that all the fields being evaluated at the beam-splitter location,
we will omit the z-dependence in the following. The difference
photocurrent is a measure of the quadrature operator:
  \begin{eqnarray}
E_{H}(\Omega)=\int_{det}\!\!\!\! d\textbf{x}
\left[B^{out}(\textbf{x},\Omega)\alpha_{L}^{*}(\textbf{x})+
B^{out+}(\textbf{x},-\Omega)\alpha_{L}(\textbf{x})\right]\,
\end{eqnarray}
where $\textit{det}$ is the reciprocal image of the photodetection
region at the beamsplitter plane, and assumed to be identical  for
the two photodetectors. We have also assumed here that the quantum
efficiency of the photodetector is equal to $1$. Here $B^{out}$ is
the sum of its odd and even part:
 \begin{eqnarray}
B^{out}(\textbf{x},\Omega)=B^{out}_{+}(\textbf{x},\Omega)
+B^{out}_{-}(\textbf{x},\Omega)
\end{eqnarray}
The fluctuations $\delta E_{H}(\Omega)$ of the homodyne field
around steady state are characterized by a noise spectrum:
\begin{eqnarray}
V(\Omega)=\int_{-\infty}^{+\infty} d\Omega'\langle\delta
E_{H}(\Omega)\delta E_{H}(\Omega')\rangle=N+S(\Omega)
\end{eqnarray}
where $E_{H}$ is normalized so that $N$ gives the mean photon
number measured by the detector
\begin{eqnarray}
N=\int_{det} dx |\alpha_{L}(x)|^{2}
\end{eqnarray}
N represents the shot-noise level, and S is the normally ordered
part of the fluctuation spectrum, which accounts for the excess or
decrease of noise with respect to the standard quantum level.

\subsection{Input/output relation}

The relation linking the outgoing fields
$B^{out}_{\pm}(\textbf{x},t)$ with the intracavity and input
fields at the cavity input/output port\cite{Gardiner} is:
 \begin{eqnarray}
B^{out}_{\pm}(\textbf{x},t)=\sqrt{2\gamma}B_{\pm}
(\textbf{x},t)-B^{in}_{\pm}(\textbf{x},t)
\end{eqnarray}
Equation (13) in the near-field (or (21) in the far field case) is
easily solved in the frequency domain, by introducing:
$$
B^{in/out}_{\pm}(\textbf{x},\Omega)=\int\frac{dt}{\sqrt{2\pi}}B^{in/out}_{\pm}(\textbf{x},t)e^{-i\Omega
t }
$$
Taking into account the boundary condition (37), we obtain the
input/output relation:
%\begin{eqnarray}
%& & [i\Omega+\gamma(1+i\Delta)][B^{out}_{+}(\textbf{x},\Omega)+B^{in}_{+}(\textbf{x},\Omega)]
% =  2\gamma B^{in}_{+}(\textbf{x},\Omega) +
% \nonumber\\
%& & \frac{\gamma}{i\Omega+\gamma(1-i\Delta)}\int\!\!\!\int
%d^{2}\textbf{x}^{'}K_{int}(\textbf{x},\textbf{x}^{'}) \Big[  2 \gamma
%B^{in+}_{+}(\textbf{x}^{'},-\Omega) \nonumber \\
%& & +\int\!\!\!\int
%d^{2}\textbf{x}^{''} \gamma
%K_{int}^{*}(\textbf{x}^{'},\textbf{x}^{''})
%(B^{in}_{+}(\textbf{x}^{''},\Omega)+B^{out}_{+}(\textbf{x}^{''},\Omega))\Big]
%\end{eqnarray}
\begin{widetext}
\begin{eqnarray}
[i\Omega+\gamma(1+i\Delta)][B^{out}_{+}(\textbf{x},\Omega)+B^{in}_{+}(\textbf{x},\Omega)]
& = & 2\gamma B^{in}_{+}(\textbf{x},\Omega) +
\frac{\gamma}{i\Omega+\gamma(1-i\Delta)}\int\!\!\!\int
d^{2}\textbf{x}^{'}K_{int}(\textbf{x},\textbf{x}^{'}) \Big[  2 \gamma
B^{in+}_{+}(\textbf{x}^{'},-\Omega) \nonumber\\
 && +\int\!\!\!\int
d^{2}\textbf{x}^{''} \gamma
K_{int}^{*}(\textbf{x}^{'},\textbf{x}^{''})
(B^{in}_{+}(\textbf{x}^{''},\Omega)+B^{out}_{+}(\textbf{x}^{''},\Omega))\Big]
\end{eqnarray}
\end{widetext}
In the case of a thin crystal in the near field\cite{Petsas} or a
plane pump in the far field, this relation describes an infinite
set of independent optical parametric oscillators. In these cases
the squeezing spectrum can be calculated analytically as we will
see in the following. But in other cases, this relation links all
points in the transverse plane. In order to get the input/output
relation, we have to inverse relation (38) by using a numerical
method.

\subsection{Numerical method}
 In order to inverse relation (38) by numerical means, we need to
discretise the transverse plane in order to replace integrals by
discrete sums. For the sake of simplicity, we will only describe
here the solution in the single transverse dimension model: the
cavity is assumed to consist of cylindrical mirrors, so that the
the transverse fields depend on a single parameter, y. In this
case the electromagnetic fields are represented by vectors and the
interaction terms by matrices. Straightforward calculations show
that we can introduce the interaction functions $U(y,y')$ and
$V(y,y')$ (calculated at resonance $\Delta=0$ and at zero
frequency in near-field or far-field configurations ) linking two
different points in the transverse plane, so that relation (38)
becomes:
\begin{eqnarray}
B_{+}^{out}(y)& =& \int_{-\infty}^{\infty}dy'U(y,y')B_{+}^{in}(y')\nonumber \\
& & + \int_{-\infty}^{\infty}dy'V(y,y')B_{+}^{in+}(y')
\end{eqnarray}

Since we assumed that the odd part of the output field is in the
vacuum state, $ B_{-}^{out}$ gives no contribution to the normally
ordered part of the spectrum $S$, which can be calculated by using
the input/output relation (21) for the even part of the field, and
by using the commutation rules for the even part:
\begin{eqnarray}
[B^{in/out}_{\pm}(x,t),B^{in/out+}_{\pm}(x',t')]=\nonumber \\
\frac{1}{2}
[\delta(x-x')\pm\delta(x+x')]\delta(t-t')
\end{eqnarray}
In the following, we will assume, as in
Refs.\cite{Petsas}\cite{Grangier}, that the local oscillator has a
constant phase profile $\varphi_{L}(\textbf{x})=\varphi_{L}$, so
that $\alpha_{L}(x)=|\alpha_{L}(x)|e^{i\varphi_{L}}$. We obtain
the ordered part of the spectrum, normalized to the shot noise:
%\begin{eqnarray}
%\frac{S(0)}{N}& = & \frac{1}{\int_{det}dy|\alpha_{L}|^{2}}
%             \int\int_{det^{2}}dx
%dx'\int_{-\infty}^{+\infty}dy|\alpha_{L}(x)||\alpha_{L}(x')|\nonumber \\
%& & \Big[(V(x,y)V(x',y)+V(x,y)V(x,-y)) \\
%& & +\cos(2\varphi_{L})(U(x,y)V(x',y)+U(x,y)V(x',-y))\Big]\nonumber
%\end{eqnarray}
\begin{widetext}
\begin{eqnarray}
\frac{S(0)}{N} = \frac{1}{\int_{det}dy|\alpha_{L}|^{2}}
             \int\int_{det^{2}}dx
dx'\int_{-\infty}^{+\infty}dy|\alpha_{L}(x)||\alpha_{L}(x')|
 &&\Big[(V(x,y)V(x',y)+V(x,y)V(x,-y)) \nonumber \\
&& +\cos(2\varphi_{L})(U(x,y)V(x',y)+U(x,y)V(x',-y))\Big]
\end{eqnarray}
\end{widetext}
Now, knowing the $U(y,y')$ and $V(y,y')$ interaction functions, we
are able to calculate the squeezing spectrum in both near and
far-field cases.

\section{Squeezing spectrum in the near-field}
\label{sec:effet}

In this section, we use the near-field homodyne detection (Fig.
\ref{fig:fig3}) described in Ref.\cite{Petsas}. As already said in
part II, in the near field, the thick crystal couples pixels
contained in a region whose size is in the order of $l_{coh}$
(18).

\begin{figure}[ht]
\centerline{
    \scalebox{.5}{\includegraphics*{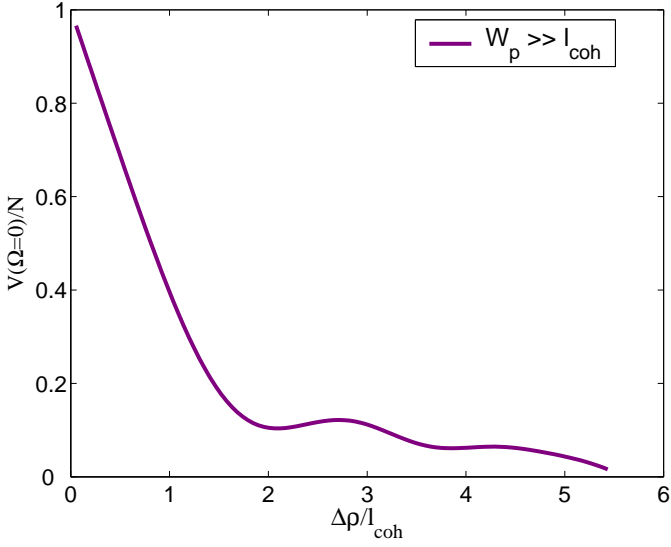}} }
\caption{ Squeezing spectrum at zero-frequency, normalized to the
shot noise, as a function of the detector radius (scaled to
$l_{coh}$).} \label{fig:fig5}
\end{figure}

Let us study first the case of a plane-wave pump and a plane-wave
local oscillator. As pointed out in ref\cite{Grangier}, in this
case and in the thin crystal approximation, the level of squeezing
does not depend on the width of the detection region. Fig.
\ref{fig:fig5} shows results predicted for a measurement performed
with a circular detector of radius $\triangle\rho$ centered on the
cavity axis (which is a symmetric detection area, as pointed out
in \cite{Petsas}). We represent the squeezing spectrum at zero
frequency as a function of the size of the detector, scaled to the
coherence length $l_{coh}=\sqrt{\frac{\lambda l_{c}}{\pi n_{s}}}
$. We can see that for $\Delta\rho<l_{c}$, the squeezing tends to
zero when $\Delta\rho\longrightarrow0$, as already predicted. For
larger values of the detector size, perfect squeezing can be
achieved. We can also see that the squeezing evolution is
comparable to the $\triangle$ function evolution (Fig.
\ref{fig:fig2}).

\begin{figure}[ht]
\centerline{
    \scalebox{.5}{\includegraphics*{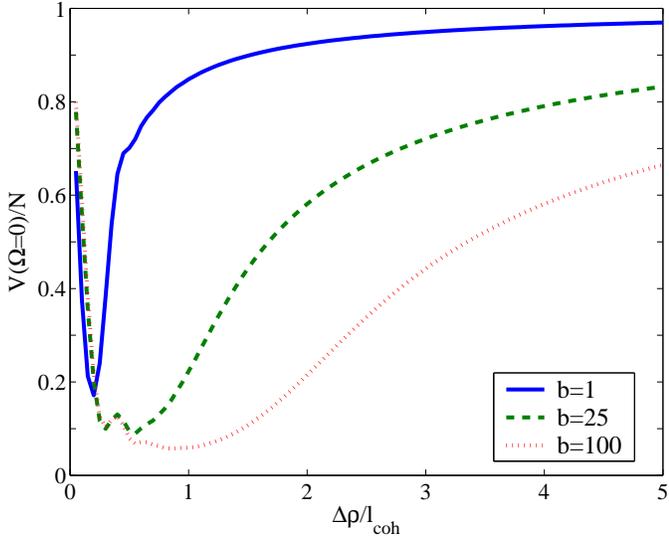}} }
\caption{Squeezing spectrum at zero-frequency, normalised to the
shot noise, as a function of the radial amplitude of the detector
scaled to $ l_{coh} $, plotted for several values of b }
\label{fig:fig6}
\end{figure}

 In the more realistic case of finite size pump, the
squeezing level depends on the parameter $b=\frac{
w_p^2}{l_{coh}^2}= 2 n_s \frac{z_p}{l_c}$, as pointed out in part
I. Fig. \ref{fig:fig6} represents the squeezing spectrum at
zero-frequency as a function the detector radius, normalized to
$l_{coh}$, for different b parameters, using a plane local
oscillator. As already seen in Fig. \ref{fig:fig5}, for
$\Delta\rho\longrightarrow0$, the noise reduction effect tends to
zero. But we see now that there is also no squeezing effect for
large values of the detector radius, because of the finite size of
the pump, as already shown in \cite{Petsas}.

\begin{figure}[ht]
\centerline{
    \scalebox{.5}{\includegraphics*{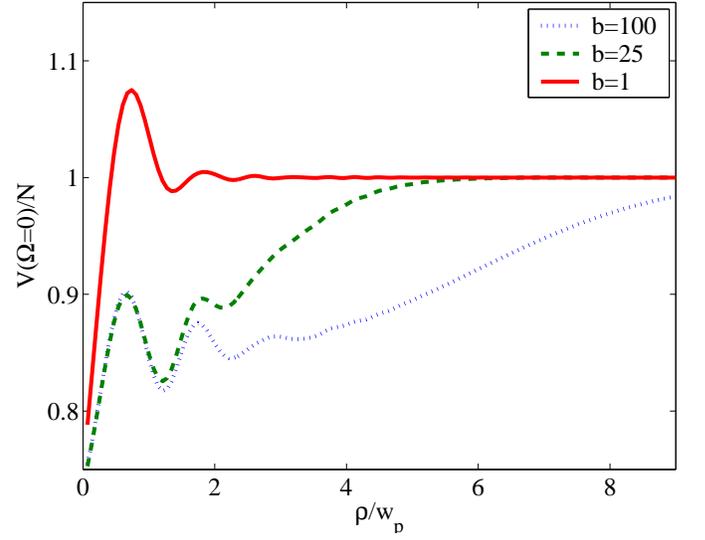}} }
\caption{Squeezing spectrum at zero-frequency, normalised to the
shot noise, as a function of the pixel distance  between the two
pixels $\rho$ from the cavity axis (scaled to $l_{coh} $), plotted
for several values of b} \label{fig:fig7}
\end{figure}

Fig. \ref{fig:fig7} shows theoretical results in the case of a
detector consisting of two symmetric pixels (pixel of size equal
to the coherence length), for different b values, in function of
the distance between the two pixels.
For large values of $\rho$, the noise level goes back to shot
noise because of the finite size of the pump, as already depicted
in reference \cite{Petsas}. But now, for small $\rho$ values, the
squeezing does not tend to zero, as in the thin crystal case.

\section{Squeezing spectrum in the far-field}
\label{sec:far}

In this section, we will consider the spatial squeezing spectrum
in the far field (Fig. \ref{fig:fig4}) and in the q-vector basis.
As already said in section II.C, the coupling between q-vectors
modes is now due to the finite length of the pump. We will see
that a new coherence length $l_{cohf}$ appears in such a case,
given by:
$$l_{cohf}\propto\frac{1}{w_{p}}$$. We will successively investigate
two configurations: the plane wave pump regime (where the
squeezing spectrum can be calculated analytically), and the case
of a finite pump size (where a numerical method is necessary).

\subsection{Plane wave pump regime in the far field}

In order to evaluate the far field case, we introduce the spatial
Fourier transforms of the electromagnetic field temporal frequency
components:
$$
\tilde{B}^{in/out}_{\pm}(\textbf{q},\Omega)=\int\int\frac{d^{2}x}{2\pi}\hat{B}^{in/out}_{\pm}(\textbf{x},\Omega)e^{-i\textbf{q}.\textbf{x}}
$$
In the case of a plane wave pump, $A_{p}(\textbf{x},z)=A_{p}$, so
that equation(22) becomes:
\begin{eqnarray}
\frac{\partial \tilde{B}_{+}}{\partial
t}(\textbf{q},0,t)&=&-\gamma[(1+i\Delta)\tilde{B}_{+}(\textbf{q},0,t)
+\sqrt{2 \gamma}\tilde{B}_{+ \, in}(\textbf{q},0,t)\nonumber\\
&&-A_{p}
sinc(\frac{l_{c}\textbf{q}^{2}}{2k_{s}})\tilde{B}_{+}^{\dagger}(\textbf{q},0,t)]
\end{eqnarray}
This equation, which does not mix different $\textbf{q}$ values,
can be solved analytically. It is similar to equation (14) in
reference \cite{Petsas}. Taking into account the boundary
condition
\begin{eqnarray}
\tilde{B}^{out}_{\pm}(\textbf{q},t)=\sqrt{2\gamma}\tilde{B}_{\pm}(\textbf{q},t)-\tilde{B}^{in}_{\pm}(\textbf{q},t)
\end{eqnarray}
 We obtain:
\begin{equation}
\tilde{B}^{out}_{+}(\textbf{q},\Omega)=\textit{U}(\textbf{q},\Omega)\tilde{B}^{in}_{+}(\textbf{q},\Omega)+\textit{V}(\textbf{q},\Omega)\tilde{B}^{in+}_{+}(-\textbf{q},-\Omega)
\end{equation}
where
\begin{equation}
\textit{U}(\textbf{q},\Omega)=\frac{[1-i(\Delta-\Omega/\gamma)][1-i(\Delta+\Omega/\gamma)]+A_{p}^{2}sinc^{2}(\frac{l_{c}\textbf{q}^{2}}{2k_{s}})}{[1+i(\Delta+\Omega/\gamma)][1-i(\Delta-\Omega/\gamma)]-A_{p}^{2}sinc^{2}(\frac{l_{c}\textbf{q}^{2}}{2k_{s}})}
\end{equation}
and
\begin{eqnarray}
&&\textit{V}(\textbf{q},\Omega)=\\
&&\frac{2
A_{p}sinc(\frac{l_{c}\textbf{q}^{2}}{2k_{s}})}{[1+i(\Delta+\Omega/\gamma)][1-i(\Delta-\Omega/\gamma)]-A_{p}^{2}sinc^{2}(\frac{l_{c}\textbf{q}^{2}}{2k_{s}})}\nonumber
\end{eqnarray}
In the case of plane wave regime, the input/output relation in the
spatial Fourier space describes therefore an infinite set of
independent optical parametric oscillators below threshold. This
can be simply understood: the q-vector basis is the eigenbasis of
the diffraction, so that no coupling between q-vector modes due to
the crystal appears.

Let us now consider the homodyne-detection scheme, schematically
shown in (Fig. \ref{fig:fig4}). The lens provides a spatial
Fourier transform of the output field $B_{out}(x,\Omega)$, so that
at the location of plane D the field $B_{out}^{D}(x,\Omega)$ is:
\begin{eqnarray}
B_{out}^{D}(x,\Omega)=\frac{2 \pi}{\lambda f}
\tilde{B}_{out}(\frac{2\pi}{\lambda f} x,\Omega)
\end{eqnarray}
In this plane,$B_{out}^{D}(x,\Omega)$ is mixed with an intense
stationary and coherent beam
$\alpha_{LO}^{D}(x)=\frac{2\pi}{\lambda f
}\tilde{\alpha}_{LO}(\frac{2 \pi x }{\lambda f},\Omega)$, where
$\alpha_{L}(x)$ has a gaussian shape, with a waist $w_{LO}$. The
homodyne field has thus an expression similar to the near field
case, where functions of $x$ are now replaced by their spatial
Fourier transforms:
\begin{eqnarray}
&&E_{H}(\Omega)= \\
&&\int_{det}d\textbf{q}
[\tilde{B}^{out}(\textbf{q},\Omega)\tilde{\alpha}_{LO}^{*}(\textbf{q})+
\tilde{B}^{out+}(\textbf{q},-\Omega)\tilde{\alpha}_{LO}(\textbf{q})]\nonumber
\end{eqnarray}
This analogy shows that, in the case of a local oscillator that
has an even parity with respect to coordinate inversion, the
squeezing spectrum is given by (like in \cite{Petsas}):
\begin{eqnarray}
V(\Omega)&=&\int_{det}d\textbf{q}\{|\tilde{\alpha}_{LO}(\textbf{q})|^2[1-\sigma(\textbf{q})]\}\nonumber\\
&&+\int_{det}d\textbf{q}\{|\tilde{\alpha}_{LO}(\textbf{q})|^2\sigma(\textbf{q})R(\textbf{q},\Omega)\}
\end{eqnarray}
where the noise spatial density $R(\textbf{q},\Omega)$ is given
by:
\begin{eqnarray}
R(\textbf{q},\Omega)=|\textit{U}(\textbf{q},\Omega)+e^{2i\varphi_{LO}(\textbf{q})}\textit{V}^{*}(\textbf{q},-\Omega)|^{2}
\end{eqnarray}
and where
\begin{eqnarray}
\sigma(\textbf{q})=\int_{det}d\textbf{q'}\delta_{+}(\textbf{q},\textbf{q'})
\end{eqnarray}

\begin{figure}[ht]
\centerline{
    \scalebox{.5}{\includegraphics*{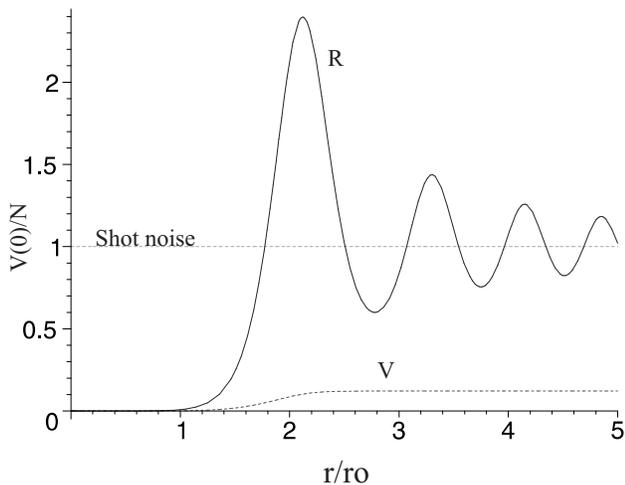}} }
\caption{Squeezing spectrum normalized to the shot noise,at
zero-frequency, at resonance, in the plane pump regime and far
field case, for two measurement configurations.  $V$ is obtained
using a circular detector of radial amplitude r (scaled to
$r_{0}$). $R$ is obtained using a pair of symmetrical pixels in
function of the pixel distance from the axis r (scaled to
$r_{0}$)} \label{fig:fig8}
\end{figure}

In order to minimize $R(\textbf{q},\Omega)$, the local oscillator
phase should be chosen as $\varphi_{LO}(\textbf{q})=
\frac{arg[U(\textbf{q},\Omega)V(\textbf{q},\Omega)]}{2}$. In
particular, at resonance and at zero frequency $U(\textbf{q},0)$
and $V(\textbf{q},0)$ are real and the optimal local oscillator
phase would correspond to
$\varphi_{LO}(\textbf{q})=\frac{\pi}{2}$, when
$sinc(\frac{l_{c}\textbf{q}^{2}}{2k_{s}})\geq0$, and
$\varphi_{LO}(\textbf{q})=0$, when
$sinc(\frac{l_{c}\textbf{q}^{2}}{2k_{s}})\leq0$, which is not
indeed very practical. However, modes for which
$sinc(\frac{l_{c}\textbf{q}^{2}}{2k_{s}})\leq0$ are quite outside
the phase matching curve, so that  the choice
$\varphi_{LO}(\textbf{q})=\frac{\pi}{2}$ everywhere should give
good results. The squeezing spectrum at resonance and zero
frequency, for $\varphi_{LO}(\textbf{q})=\frac{\pi}{2}$ can be
analytically calculated and, as a function of the radius $r$  of a
detector centered on the optical axis is given by:
\begin{eqnarray}
\frac{V(r,0)}{N}&=&\frac{1}{\int_{0}^{r/r_{0}}u
\exp(\frac{-w_{LO}^{2}k_{s}u^{2}}{l_{c}})}\\
&&*\int_{0}^{r/r_{o}}\!\!\!\!\! u
\exp\left(\frac{-w_{LO}^{2}k_{s}u^{2}}{l_{c}}\right)
\left(\frac{1+A_{p}sinc(u^{2})}{1-A_{p}sinc(u^{2})}\right)^{2}\nonumber
%\int_{0}^{q/q_{0}}r
%\exp(\frac{-w_{LO}^{2}k_{s}r^{2}}{l_{c}})
\end{eqnarray}
where
\begin{eqnarray}
r_{0}=\frac{\lambda f}{2
\pi}\sqrt{\frac{2k_{s}}{l_{c}}}=\frac{\lambda
f}{\pi}*\frac{1}{l_{coh}}
\end{eqnarray}
Fig. \ref{fig:fig8} shows the results obtained in the case of two
different detection configurations: the $V$ curve shows results in
the case of a circular detector of variable radius $r$ (scaled to
$r_{0}$), using a local oscillator waist $w_{LO}=r_{0}$. As
already said in part $II$, the limitation of the squeezing level
is due to the non perfect phase matching along the crystal. For
$r>r_{0}$, the squeezing level decreases. So, in the plane wave
pump regime in the far field, the thickness of the crystal has a
role comparable with the finite size of the pump in the near
field, as reported in \cite{Petsas}. The $R$ curve shows results
obtained in the case of two small symmetrical pixels and a plane
wave local oscillator as a function of the pixel distance from the
cavity axis $r$, scaled to $r_{0}$. We can see that the noise
level goes back to the shot noise level for $r>r_{0}$, because to
the non perfect phase matching along the crystal.

\subsection{Squeezing spectrum in the far field case and finite size pump regime}

When one takes into account the finite size of the pump, a
coupling between different q vectors appear, and one needs to
solve equations numerically, as in the near field case. A new
coherence length $l_{cohf}$ appears in the far field:
$l_{cohf}=\frac{1}{w_{P}}$.

\begin{figure}[ht]
\centerline{
    \scalebox{.5}{\includegraphics*{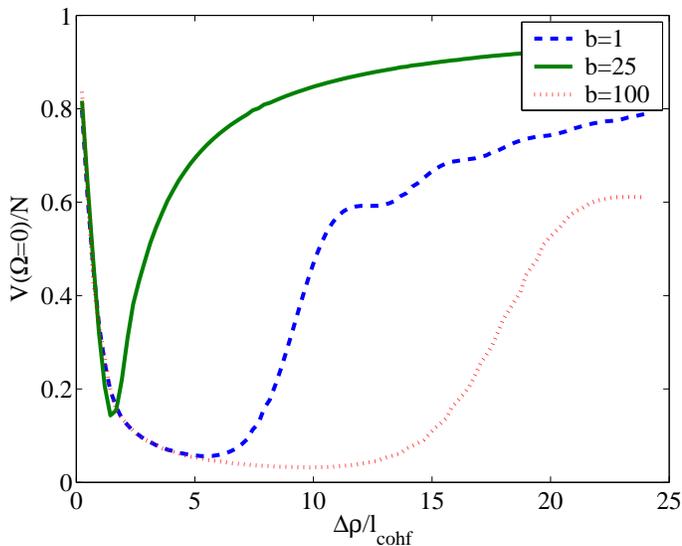}} }
\caption{Squeezing spectrum normalized to the shot noise at
zero-frequency, and at resonance, as a function of the radial
amplitude of the detector $\Delta\rho$(scaled to the coherence
area $l_{cohf}$), in the finite pump regime and far field approach
and for different values of b.} \label{fig:fig9}
\end{figure}

Fig. \ref{fig:fig9} shows the evolution of the squeezing spectrum
at zero frequency, and at resonance, for different b parameters,
in function of the detector radius scaled to $l_{cohf}$ . We see
the same evolution as in the analytical case, except that the
noise level tends to shot noise for small values of the detector.

\begin{figure}[ht]
\centerline{
    \scalebox{.5}{\includegraphics*{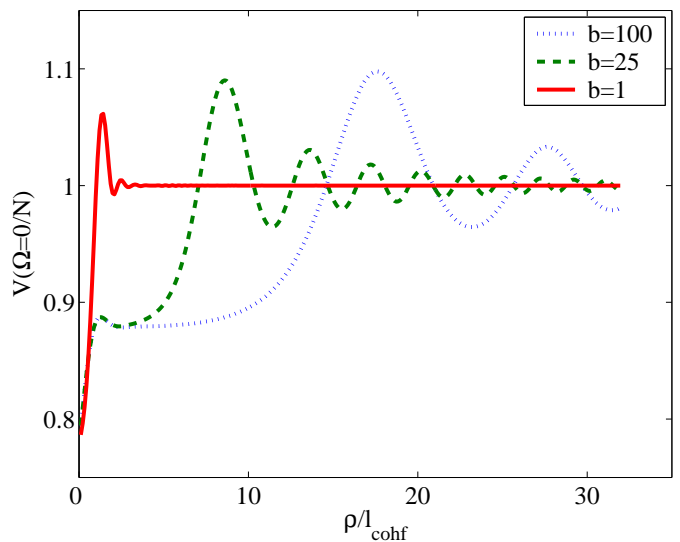}} }
\caption{Squeezing spectrum normalized to the shot noise at
zero-frequency, and at resonance,
 as a function of the distance between the two pixels  $\rho$ (scaled to the coherence area $l_{cohf}$), in the finite pump
regime and far field approach and for different values of b.}
\label{fig:fig10}
\end{figure}

Fig. \ref{fig:fig10} shows the results obtained in the case of two
 symmetrical pixels (pixel of size equal to the coherence
length $l_{cohf}$), for different b values, in function of the
distance between the two pixels $\rho$. The evolution is similar
to the one given by fig(6) for large distances. But there is here
also a decrease of the squeezing effect for small distances.

\section{Discussions and Conclusions}
\label{sec:conclusion}

We have seen that when one takes into account the effect of
diffraction inside the nonlinear crystal in a confocal OPO, the
local squeezing predicted for any shape and size of the detectors
in the thin crystal approximation is now restricted to detection
areas lying within a given range, characterized by a coherence
length $l_{coh}$. This prediction introduces serious limitations
to the success of an experiment, and must be taken into account
when designing the experimental set-up. With the purpose of
producing a light beam that is squeezed in several elementary
portions of its transverse cross-section, either a crystal short
compared to $z_R$ should be used or, alternatively, a defocussed
pump, with a waist much larger than the cavity waist. In both
cases the efficiency of the non linear coupling is reduced. For
instance, with $1cm$ long crystal, $l_{coh}$ is equal to $40\mu
m$, and one must choose a pump waist much larger than this value
in order to observe multimode squeezing (the number of modes being
roughly equal to the ratio $b=\frac{w_{p}^{2}}{l_{coh}^{2}}$).
This defocussed pump will imply a much higher threshold for the
OPO oscillation, which is multiplied by a factor also close to
$b$. The conclusion of this analysis is that one cannot have
multimode squeezing "for free", and that with a given pump power,
one will be able to to excite a number of modes which is roughly
equal to the ration of the injected pump power to the threshold
power for single mode operation.

\begin{acknowledgments}
Laboratoire Kastler-Brossel, of the Ecole Normale Sup\'{e}rieure
and the Universit\'{e} Pierre et Marie Curie, is associated with
the Centre National de la Recherche Scientifique.\\
This work was supported by the European Commission in the frame of
the QUANTIM project (IST-2000-26019).\\
\end{acknowledgments}

\end{document}